\documentclass[floats,aps,onecolumn,showpacs]{revtex4}
\usepackage{dcolumn}
\usepackage{epsfig}

\newcommand {\beq} {\begin{eqnarray}}
\newcommand {\eeqn} [1] {\label{#1} \end{eqnarray}}%

\begin{document}
%\initfloatingfigs
%\tighten
\title{Relativistic structure of one-meson and one-gluon exchange forces
 and the lower excitation spectrum of the Nucleon and the $\Delta$}

\author{
E.\ M.\ Tursunov
%$^{1)}$
}
\affiliation{
%$^{1)}$
 Institute of Nuclear Physics, Uzbekistan
Academy of Sciences, 100214, Ulugbek, Tashkent, Uzbekistan
 }

\date{\today}
\begin{abstract}
The lower excitation spectrum of the nucleon and $\Delta$ is
calculated in a relativistic chiral quark model. Corrections to the
baryon mass spectrum from the second order self-energy and exchange
diagrams induced by pion and gluon fields are estimated in the field
-theoretical framework. Convergent results for the self-energy terms
are obtained when including the intermediate quark and antiquark
states with a total momentum up to $j=25/2$. Relativistic one-meson
and color-magnetic one-gluon exchange forces are shown to generate
spin 0, 1, 2, etc. operators, which couple the lower and the upper
components of the two interacting valence quarks and yield
reasonable matrix elements for the lower excitation spectrum of the
Nucleon and Delta. The only contribution to the ground state nucleon
and $\Delta$ comes from the spin 1 operators, which correspond to
the exchanged pion or gluon in the l=1 orbit, thus indicating, that
the both pion exchange and color-magnetic gluon exchange forces can
contribute to the spin of baryons. Is is shown also that the
contribution of the color-electric component of the gluon fields to
the baryon spectrum is enormously large (more than 500 MeV with a
value $\alpha_s=0.65$ ) and one needs to restrict to very small
values of the strong coupling constant or to exclude completely the
gluon-loop corrections to the baryon spectrum. With this
restriction, the calculated spectrum reproduces the main properties
of the data, however needs further contribution from the two-pion
exchange and instanton induced exchange (for the nucleon sector)
forces in consistence with the realistic NN-interaction models.
\end{abstract}

% \Keywords{Relativistic quark model; meson exchange; gluon exchange;
%   spectroscopy of N^* and $\Delta^*$ resonances}

\maketitle

\section{Introduction}
\par The spectroscopy is a serious task for any model of hadron structure. It can yield
a detailed information on the source of hyperfine interaction, as
well as on effective degrees of freedom in the description of hadron
dynamics. The so-called "three-body spin-orbit puzzle" has a long
history. In the original papers of Isgur and Karl \cite{isg77,isg78}
it was shown that the whole SU(3) baryon spectrum can be reproduced
qualitatively and (with some restrictions) quantitatively in a
constituent quark model based on a non-relativistic hamiltonian with
a central confining potential plus effective color-magnetic
one-gluon exchange (OGE) forces. The hyperfine interaction was
assumed to be a sum of the spin-spin, tensor and spin-orbit
interaction potentials between the constituent quarks. However, the
spin-orbit term yields a very large (of order 500 MeV) matrix
elements in the three-body system (typical baryons), while in mesons
and meson-like baryons \cite{isg99} the matrix elements are small as
a result of the strong cancelation between the dynamical one-gluon
exchange spin-orbit forces and a pure kinematical spin-orbit forces
due-to the Thomas precession of the quark spin in the central
confining potential. Contrary, in ordinary baryons (like $N^*$ and
$\Delta^*$), there are not enough cancelation effects between the
two origin spin-orbit forces. Showing these possibilities and
assuming that in baryons like in mesons must be a consistent
cancelation, Isgur and Karl, as a first approximation, excluded all
spin-orbit forces from the hyperfine interaction potential. Since
the beginning the problem was called a "three body spin-orbit
puzzle" in baryons and up to now it was not resolved in a consistent
way, although in the relativized version of the constituent quark
models (see \cite{cap86,sta90,sta91}) some reduction of the size of
the spin-orbit interactions to acceptable levels was obtained.
\par There are other strong candidates for the hyperfine interaction in baryons such as
Goldstone-boson exchange (GBE) forces \cite{glo96,glo98} and the
instanton induced exchange forces (IIE) \cite{met01}. The GBE based
constituent quark model was proposed as a realization of the
spontaneous breaking of the chiral symmetry of QCD, which plays a
crucial role for the hadron structure. The debates between the OGE
and GBE based quark models over the last decade
\cite{glo97,isg99,cap00} were focused on the "three-body spin-orbit
puzzle", as well as the level ordering problem and the mixing angles
in excited baryon sector. While reproducing the correct level
ordering (for example $N^*(1/2^+)(1440)$ - $N^*(3/2^-)(1520)$,
$N^*(1/2^-)(1535)$ ), the GBE based model does not contain
 spin-orbit terms.

\par The main issue of these debates likely supports the idea that the
origin of hyperfine interaction between valence quarks is a sum of
the GBE and OGE forces (and possible  two-pion exchange
\cite{ris99}, and additionally IIE forces for the nucleon sector) in
consistence with the theory of the low-energy NN-interaction
\cite{hol92}. The question is, however, how much the each exchange
mechanism does contribute to the baryon spectrum? The solving of
this complex problem requires many efforts to be done both
theoretically and experimentally.

\par On the other hand, from the beginning of the proton spin crisis \cite{EMC88} it
became clear that the non-relativistic constituent quark models are
not adequate and complete picture for the hadron structure, since
only about 30 percent of the proton spin is described by the valence
quarks! The possible contribution comes from the coupling to the
lower component of the Dirac spinor, one-gluon exchange and pion
cloud effects. The problem can be resolved in the frame of the
cloudy bag model (CBM) \cite{thomas81}, as was claimed in the recent
papers \cite{thomas08}. A relativistic picture of the hadron
structure, while playing a decisive role for the understanding of
the proton spin crisis, also should yield a consistent description
of the hadron spectrum including excited baryons. The questions,
concerning the source of the hyperfine
 splitting in baryons, the relativistic structure of the one-meson and one-gluon exchange
mechanisms, the role of the lower component of the Dirac spinor of
valence quarks and the contribution of the sea-quarks to baryon
spectrum are still being open! The use of the bag models in baryon
spectroscopy was restricted to the octet and decuplet baryons
\cite{saito84} while having some specific problems concerning the
convergence of the self-energy and the sharp surface pion-quark
coupling. The recent progress in the frame of the chiral quark-meson
coupling model (CQMCM)\cite{saito08} was done for the description of
the N and $\Delta$ spectrum and the Nuclear Matter with inclusion
both gluon and meson one-loop diagrams while restricting the
intermediate baryon states to the lowest mode when considering the
self-energy diagrams.

\par The aim of present paper is to show that a description of the baryon structure in a
field-theoretical based chiral quark model
\cite{oset84,gutsche87,gutsche89,tur05} can open many windows to the
long-standing problems of excited baryon spectroscopy. In the last
work we have calculated the lower excitation spectrum of the SU(2)
baryons on the basis of one pion-loop diagrams. Further we will show
that relativistic one-meson  and color-magnetic one-gluon exchange
forces between valence quarks generate spin 0, spin 1, spin 2 etc.
operators, which can be considered as analogy of the
non-relativistic spin-spin, spin-orbit, tensor etc. operators,
respectively. However, all these relativistic operators couple the
upper and the lower Dirac components of the two interacting valence
quarks, hence have a strongly different nature. For instance, the
two valence quarks in S-waves, interacting via one-gluon or one pion
exchange forces are coupled only by the spin 1 operators (the
analogy of the spin-orbit forces), while in the non-relativistic
models the corresponding terms have a spin-spin structure, i.e. are
spin 0 operators. This finding has an important consequence for the
proton spin problem, since the possible contribution to the proton
spin comes not only from one-gluon exchange, but also from one-meson
exchange mechanism. Additionally, in  recent paper \cite{santo08}
meson-cloud effects were shown to be important for the description
of the flavor asymmetry of the nucleon sea. We will show in a
phenomenological way that a reasonable description of the lower
excitation spectrum of the $N^*$ and $\Delta^* $ can be obtained in
a relativistic chiral quark model, including all second-order
corrections induced by pion fields (corresponding to the self energy
and exchange diagrams) to the zeroth-order quark-core results with
further possible improvement by including the two-pion exchange and
IIE (for the nucleon sector) forces. It will be shown also that the
contribution of the color-electric (Coulomb) component of the gluon
fields to the baryon spectrum is enormously large at reasonable
values of the strong coupling constant. To have an acceptable
contribution of the gluon one-loop diagrams one needs a small value
of the strong coupling constant which results the gluon-exchange
contribution to the nucleon-$\Delta$ mass splitting to be
negligible. This result confirm  the prediction of the
non-relativistic constituent quark model \cite{glo97}, that the
one-gluon exchange mechanism does not play an important role in the
description of the SU(2) baryon spectrum. We will show also that the
restriction of the intermediate baryon states to the lowest mode
when calculating the contributions of the self-energy diagrams is
 not a good approximation and in order to have convergent results,
 one has to include all intermediate excited  quark and antiquark
 states  with the total angular momentum up to j=25/2.

\par The relativistic quark model is based on an effective chiral Lagrangian
describing quarks as relativistic fermions moving in a confining
static potential. The modification of the model, the so-called
Perturbative Chiral Quark model recently was applied to the ground-
state baryon masses \cite{inoue04}. The potential is described by a
Lorentz scalar and the time component of a vector potential, where
the latter term is responsible for short-range fluctuations of the
gluon field configurations \cite{lus81}. The model potential defines
unperturbed wave functions of the quarks which are subsequently used
in the calculations of baryon properties. The baryons are considered
as bound states of valence quarks, surrounded by a pion cloud as
required by the chiral symmetry and by gluons. Interaction of quarks
with a pion is introduced on the basis of the linearized
$\sigma$-model \cite{gell60,gutsche87,gutsche89}. The residual
color-magnetic quark-gluon interaction is introduced on the
field-theoretical basis as prescribed by QCD. Calculations are
performed perturbatively to second order in the quark-pion and
quark-gluon interaction. All calculations are performed at one loop
or at order of accuracy $o(1/f_{\pi}^2, \alpha_s)$.

\par In the following we proceed as follows: we
first describe the basic formalism of our approach. Then we indicate
the main derivations relevant to the problem and finally present
the numerical results.

\section{Model}
\par The effective Lagrangian of our model
${\cal L}(x)$ contains the quark core part ${\cal L}_Q(x)$ the quark-pion
 ${\cal L}_I^{(q\pi)}(x)$ and the quark-gluon ${\cal L}_I^{(qg)}(x) $
 interaction parts, and the kinetic parts for the pion ${\cal L}_{\pi}(x)$ and
gluon ${\cal L}_{g}(x)$:
\begin{eqnarray}
\nonumber
{\cal L}(x) = {\cal L}_Q(x) + {\cal L}_I^{(q\pi)}(x) + {\cal
  L}_I^{(qg)}(x)+ {\cal L}_{\pi}(x) +  {\cal L}_{g}(x)  \\
 \nonumber
 = \bar\psi(x)[i\not\!\partial -S(r)-\gamma^0V(r)]\psi(x) - 1/f_{\pi}
 \bar\psi[S(r) i \gamma^5 \tau^i \phi_i]\psi- \\
 -g_s \bar\psi A_{\mu}^a\gamma^{\mu}\frac{\lambda^a}{2} \psi +
  \frac{1}{2}(\!\partial_{\mu}\phi_i)^2 -{1\over 4}
G^a_{\mu\nu} G_a^{\mu\nu}.
\end{eqnarray}
Here, $\psi(x)$, $\phi_i, i=1,2,3$ and $A_{\mu}^a$ are the quark,
the pion and the gluon fields, respectively. The matrices $\tau^i
(i=1,2,3)$ and $\lambda^a (a=1,...,8)$ are the isospin and color
matrices, correspondingly. The pion decay constant $f_\pi=$93 MeV.
The scalar part of the static confinement potential is given by
\begin{equation}
S(r)=cr+m
\end{equation}
where c and m are constants.

\par At short distances, transverse fluctuations of the string are dominating
\cite{lus81}, with some indication that they transform like the time component of the
Lorentz vector. They are given by a Coulomb type vector potential as
\begin{equation}
\label{Coulomb}
 V(r)=-\alpha/r
\end{equation}
where $\alpha$ is approximated by a constant.
The quark fields are obtained from solving the Dirac equation with the corresponding
scalar plus vector potentials
\begin{equation} \label{Dirac}
[i\gamma^{\mu}\partial_{\mu} -S(r)-\gamma^0V(r)]\psi(x)=0
\end{equation}
The respective positive and negative energy eigenstates as solutions to the Dirac
equation with a spherically symmetric mean field, are given in a general form as
\begin{eqnarray} \label{Gaussian_Ansatz}
 u_{\alpha}(x) \, = \,
\left(
\begin{array}{c}
g^+_{N\kappa }(r) \\
-i f^+_{N\kappa }(r) \,\vec{\sigma}\hat{\vec x} \\
\end{array}
\right)
\, {\cal Y}_{\kappa}^{m_j}(\hat{\vec x}) \,\chi_{m_t} \, \chi_{m_c} \, exp(-iE_{\alpha}t)
\end{eqnarray}

\begin{eqnarray}
 v_{\beta}(x) \, = \,
\left(
\begin{array}{c}
g^-_{N\kappa}(r) \\
-i f^-_{N\kappa}(r) \,\vec{\sigma}\hat{\vec x} \\
\end{array}
\right)
\, {\cal Y}_{\kappa}^{m_j}(\hat{\vec x}) \,\chi_{m_t} \, \chi_{m_c} \, exp(+iE_{\beta}t)
\end{eqnarray}
The quark and anti-quark eigenstates $u$ and $v$ are labeled by the
radial, angular, azimuthal, isospin and color quantum numbers $N,\,
\kappa,\, m_j,\, m_t$ and $m_c$, which are collectively denoted by
$\alpha$ and $\beta$, respectively. The spin-angular part of the
quark field operators
\begin{equation}
{\cal Y}_{\kappa}^{m_j}(\hat{\vec x})\,=\,[Y_l(\hat{\vec x})\otimes
\chi_{1/2}]_{jm_j} \, \, j=|\kappa|-1/2.
\end{equation}
The quark fields $\psi$ are expanded over the basis of positive and negative energy
eigenstates as
\begin{equation}
\psi(x)=\sum \limits_{\alpha} u_{\alpha}(x)b_{\alpha} +\sum \limits_{\beta} v_{\beta}(x)d^{\dag}_{\beta} .
\end{equation}
The expansion coefficients $b_{\alpha}$ and $d^{\dag}_{\beta}$ are
operators, which annihilate a quark and create an anti-quark in the
orbits $\alpha$ and $\beta$, respectively.
\par The free pion field operator is expanded over plane wave solutions as
\begin{equation}
\phi_j(x)=(2\pi)^{-3/2}\, \int\frac{d^3k}{(2\omega_k)^{1/2}}[a_{j{\bf k}}exp(-ikx)+a^{\dag}_{j{\bf k}}exp(ikx)]
\end{equation}
with the usual destruction and creation operators $a_{j{\bf k}}$ and
$a^{\dag}_{j{\bf k}}$ respectively. The pion energy is defined as \\
$\omega_k \,=\, \sqrt{k^2+m_{\pi}^2}. $ The expansion of the free
zero mass gluon field operators is of the same form.
\par In denoting the three-quark vacuum state by $ |0> $, the corresponding
noninteracting many-body quark Green's function (propagator) is given by
the customary vacuum Feynman propagator for a binding potential \cite{fet71}:
\begin{equation}
iG(x,x')\,=\, iG^F(x,x')\,=\,<0|T\{\psi(x) \bar\psi(x')\}|0>\,=\,
\sum \limits_{\alpha} u_{\alpha}(x)\bar u_{\alpha}(x')\theta(t-t') +
\sum \limits_{\beta} v_{\beta}(x)\bar v_{\beta}(x')\theta(t'-t)
\end{equation}
Since the three-quark vacuum state $|0>$ does not contain any pion
or gluon, the pion and gluon Green's functions are given by the
usual free Feynman propagator for a boson field:
\begin{equation}
i\Delta_{ij}(x-x')\,=\, <0|T\{\phi_i(x) \bar\phi_j(x')\}|0>\,=\,
i\delta_{ij}\int\frac{d^4k}{(2\pi)^4}\frac{1}{k^2-m_{\pi}^2+i\epsilon}
\,exp[-ik(x-x')] \, ,
\end{equation}

\begin{equation}
i\Delta^{(\mu\nu)}_{ab}(x-x')\,=\, <0|T\{ A^a_\mu (x)
A^b_{\nu}(x')\}|0>\,=\,
 i\delta_{ab}g^{\mu \nu}\int\frac{d^4k}{(2\pi)^4}\frac{1}{k^2+i\epsilon}
\,exp[-ik(x-x')] \,,
\end{equation}
(in the Coulomb gauge), where we choose $g^{\mu\nu}=\delta_{\mu\nu}
g^{\mu\mu}$, $g^{00}=-g^{11}=-g^{22}=-g^{33}=1$ .
\par Using the effective Lagrangian and the time-ordered perturbation theory
 one can develop a calculation scheme for the lower
excitation spectrum of the nucleon and delta. In the model the quark
core result ($E_Q$) is obtained by solving Eq.(\ref{Dirac}) for the
single quark system numerically. Since we work in the independent
particle model and limited with the lower excitation spectrum of the
nucleon and Delta, the bare three-quark state of the $SU(2)$-flavor
baryons corresponds to the  structure $(1S_{1/2})^2(nlj)$ in the
non-relativistic spectroscopic notation. The corresponding quark
core energy is evaluated as the sum of single quark energies with:
$$ E_Q=2E(1S_{1/2}) + E(nlj)$$

\par The result for $E_Q$ still contains the contribution of the center of mass motion.
To remove this additional term we resort to three approximate methods, which
correct for the center of mass motion:
the $R=0$ \cite{lu98}, $P=0$ \cite{teg82} and LHO \cite{wil89} methods.
 \par The second order perturbative corrections to the energy spectrum of the
 SU(2) baryons due to the pion field ($\Delta E^{(\pi)}$) and the gluon fields ( $\Delta
 E^{(g)}$) are calculated on the basis of the Gell-Mann and Low theorem :
 \begin{eqnarray}\label{Energy_shift}
\hspace*{-.8cm}
\Delta E=<\Phi_0| \, \sum\limits_{i=1}^{\infty} \frac{(-i)^n}{n!} \,
\int \, i\delta(t_1) \, d^4x_1 \ldots d^4x_n \,
T[{\cal H}_I(x_1) \ldots {\cal H}_I(x_n)] \, |\Phi_0>_{c}
\end{eqnarray}
with $n=2$, where the relevant quark-pion and quark-gluon interaction
Hamiltonian densities are
\begin{eqnarray}
{\cal H}_I^{(q\pi)}(x)= \frac{i}{f_{\pi}}\bar\psi(x)\gamma^5
\vec\tau\vec\phi(x)S(r)\psi(x),
\end{eqnarray}
\begin{eqnarray}
{\cal H}_I^{(qg)}(x)= g_s\bar\psi(x)A_{\mu}^a(x)\gamma^{\mu}
\frac{\lambda^a}{2}\psi(x)
\end{eqnarray}
The stationary bare three-quark state $|\Phi_0>$ is constructed from
the vacuum state using the usual creation operators:
\begin{equation}
|\Phi_0>_{\alpha\beta\gamma}=b_{\alpha}^+b_{\beta}^+b_{\gamma}^+|0>,
\end{equation}
where $\alpha, \beta$ and $ \gamma$ represent the quantum numbers of
the single quark states, which are coupled to the respective baryon
configuration. The energy shift of Eq.(\ref{Energy_shift}) is
evaluated up to second order in the quark-pion and quark-gluon
interaction, and generates self-energy and exchange diagrams
contributions.

\subsection{ Self-energy diagrams contribution}
\par The self-energy terms contain contribution both from
intermediate quark $(E>0)$ and anti-quark $(E<0)$ states. These
diagrams correspond to the case when a pion or gluon is emitted and
absorbed by the same valence quark which is excited to the
intermediate quark and anti-quark states. In this way one can
estimate the contribution of the sea-quarks to the hadron spectrum
that can not be done in non-relativistic quark models.
\par
\begin{figure}[tbh]
\begin{center}
\includegraphics[width=15cm]{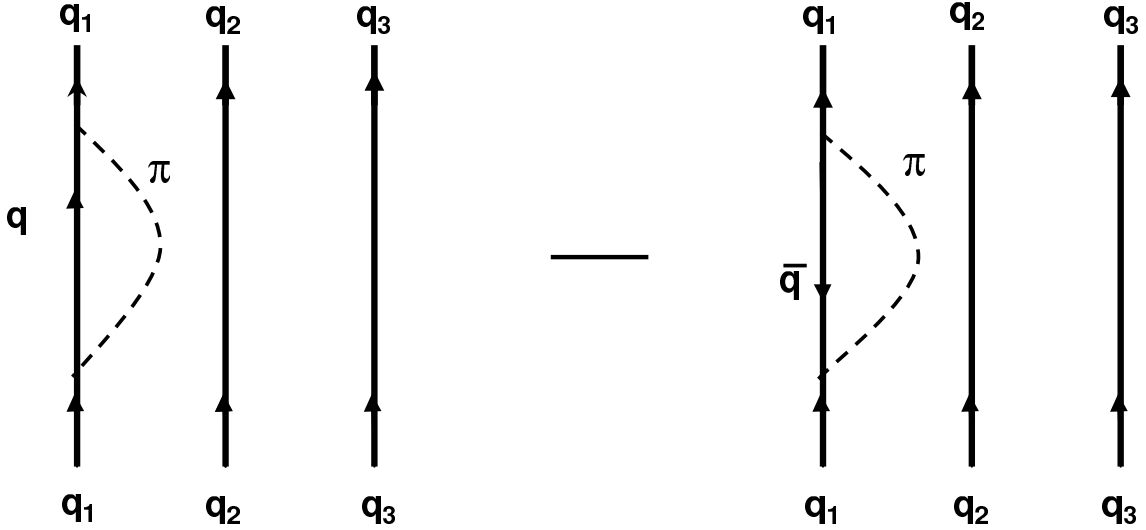}
\end{center}
\caption{Second order self energy diagrams induced by $\pi-$meson
fields \label{Fig1a}}
\end{figure}

 The pion part of the self energy term (pion cloud contribution)
 (see Fig.\ref{Fig1a} ) is evaluated as
\begin{eqnarray}
\Delta
E_{s.e.}^{(\pi)}=-\frac{1}{2f_{\pi}^2}\sum\limits_{a=1}^{3}\sum\limits_{\alpha
  ' \leq \alpha_F} \int
\frac{d^3\vec p}{(2\pi)^3p_0} \biggl\{ \sum\limits_{\alpha}\frac{V_{\alpha
      \alpha ' }^{a+}(\vec p)V_{\alpha \alpha ' }^{a}(\vec
        p)}{E_{\alpha}-E_{\alpha '}+p_0}-
 \sum\limits_{\beta}\frac{V_{\beta \alpha '}^{a+}(\vec p) V_{\beta
 \alpha '}^{a}(\vec p)}{E_{\beta}+E_{\alpha '}+p_0}\biggr\},
\end{eqnarray}
with $ p_0^2=\vec p^2 + m_{\pi}^2$. The $q-q-\pi$ transition form
factors are defined as:
\begin{eqnarray}
V_{\alpha\alpha'}^a(\vec p)=\int d^3 x \bar u_{\alpha}(\vec x)\Gamma^a(\vec x)
u_{\alpha '}(\vec x) e^{-i\vec p \vec x} \\
V_{\beta \alpha'}^a(\vec p)=\int d^3 x \bar v_{\beta}(\vec x)\Gamma^a(\vec x)
u_{\alpha '}(\vec x) e^{-i\vec p \vec x}
\end{eqnarray}
%% Dirac Hamiltonian of the single quark states with
%%\begin{eqnarray}
%%{\cal H}=\vec{\alpha}\vec p + \beta S(r) + V(r).
%%\end{eqnarray}
The vertex function of the $\pi -q -q $ and $\pi -q -\bar q $
transition is
\begin{eqnarray}
\Gamma^a= S(r) \gamma^5 \tau^a I_c \, ,
\end{eqnarray}
where $I_c$ is the color unity matrix. The estimations of the
$\pi-q-q$ and $g-q-q$ transition form factors are given in the
Appendix A and B, respectively.

After integration over the angular part in Eq. (17), the self-energy
diagrams contribution to the baryon spectrum induced by pion fields
(pion cloud contribution) is evaluated as:
\begin{eqnarray}
\nonumber \Delta E_{s.e.}^{(\pi)}=-\frac{1}{16\pi^3f_{\pi}^2}\int
\frac{d p \, p^2}{p_0} \sum\limits_{\alpha ' \leq
\alpha_F}\sum\limits_{l_n} \biggl\{ \sum\limits_{\alpha}\frac{[\int
dr r^2 G_{\alpha \alpha
    '}(r)S(r)j_{l_n}(pr)]^2}{E_{\alpha}-E_{\alpha '}+p_0}
Q_{s.e.}(l,l',l_n,j,j') - \\
\sum\limits_{\beta}\frac{[\int dr r^2 G_{\beta \alpha
'}(r)S(r)j_{l_n}(pr)]^2}{E_{\beta}+E_{\alpha '}+p_0}
Q_{s.e.}(l,l',l_n,j,j') \biggr \} ,
\end{eqnarray}
where $j_{l_n}$ is the Bessel function. The radial overlap of the
single quark states with quantum numbers
$\alpha=(N,l,j,m_j,m_t,m_c)$ and $\alpha^{\prime}$ is defined as
\begin{eqnarray}
  G_{\alpha \alpha '}(r)=f_{\alpha}(r)g_{\alpha '}(r) + f_{\alpha '}(r)
g_{\alpha }(r)  .
\end{eqnarray}
The angular momentum coefficients $Q$ are evaluated for all SU(2)
baryons as
\begin{eqnarray}
Q_{s.e.}(l,l',l_n,j,j')= 12\pi [l^{\pm}][l_n][j] \biggl [
C^{l'0}_{l^{\pm}0l_n 0} W (j \frac{1}{2}l_nl'; l^{\pm}j')\biggr ]^2
\sum\limits_{m_j}\sum \limits_{m_j ' \leq \alpha_f} \biggl [
C^{j'm_j'}_{jm_jl_n(m_j'-m_j)} \biggr ]^2 ,
\end{eqnarray}
where $C$ and $W$ are the Clebsch-Gordan and Wigner coefficients,
respectively.

\par
\begin{figure}[tbh]
\begin{center}
\includegraphics[width=15cm]{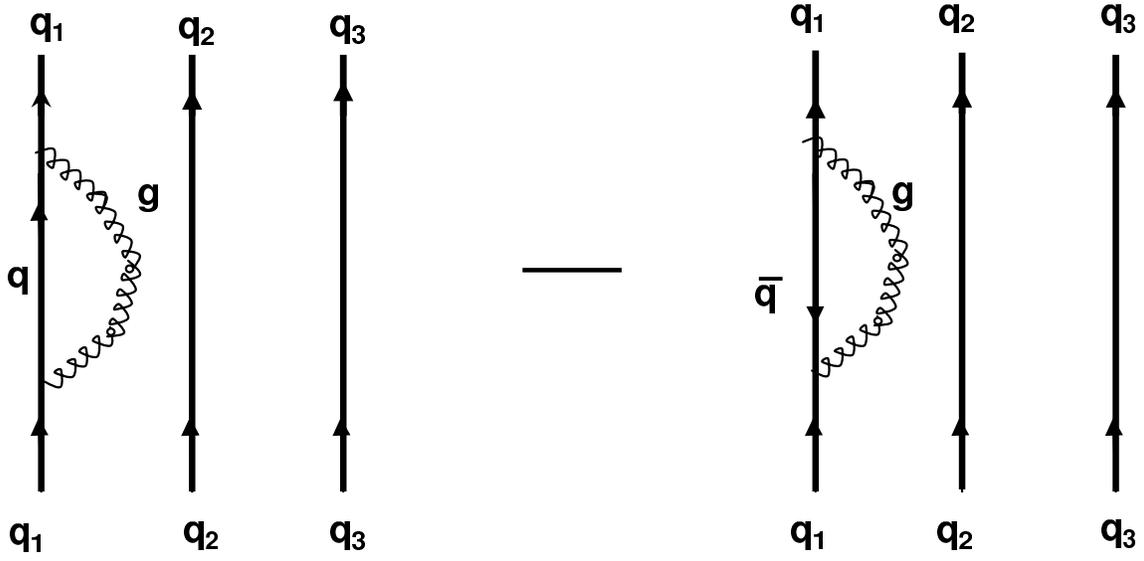}
\end{center}
\caption{Second order self energy diagrams induced by gluon fields
\label{Fig1b}}
\end{figure}

 The gluon part of the second order self-energy diagrams (gluon cloud) contribution
is  estimated in a similar fashion as
\begin{eqnarray}
\Delta E_{s.e.}^{(g)}=\frac{g_s^2}{2}\sum\limits_{a}g_{\mu\nu}
\sum\limits_{\alpha ' \leq \alpha_F} \int\frac{d^3\vec p}{(2\pi)^3p}
\biggl\{ \sum\limits_{\alpha}\frac{V_{\alpha
      \alpha ' }^{a\mu+}(\vec p)V_{\alpha \alpha ' }^{a\nu}(\vec p)}
{E_{\alpha}-E_{\alpha '}+p}-
 \sum\limits_{\beta}\frac{V_{\beta \alpha '}^{a\mu+}(\vec p) V_{\beta
 \alpha '}^{a\nu}(\vec p)}{E_{\beta}+E_{\alpha '}+p}\biggr\},
\end{eqnarray}
where the transition form factor is evaluated with the corresponding
vertex matrix
\begin{eqnarray}
\Gamma_{\mu}^a=\gamma^{\mu}\frac{\lambda^a}{2}I_t
\end{eqnarray}
with the isospin unity matrix $I_t$. The corresponding Feynman
diagrams are given in Fig.\ref{Fig1b}, where the contribution from
intermediate quark and anti-quark levels have opposite signs.
\par  After evaluation of the transition form-factors and integration over angular variables,
the self-energy term induced by gluon fields can be written as a sum
of color-electric (Coulomb) and color-magnetic parts:
%\begin{eqnarray}
\[
 \Delta E_{s.e.}^{(g)}=\frac{g_s^2}{3\pi^2} \sum\limits_{N'l'j'}\sum\limits_{(\alpha,\beta)}
\sum\limits_{LL'L^*L'^*l_n}[l_n]\left ( \frac{[L][L^*]}{[L'][L'^*]}
\right )^{1/2}C^{L'0}_{L0l_n0} C^{L'^*0}_{L^*0l_n0}
%\begin{eqnarray}
\]
\[
\biggl \{ \delta_{lLL^*}\delta_{l'L'L'^*}\delta_{l_nl}
{\cal A} ^{jj'm_jm_j'}_{LL'L^*L'^*l_n}  \\
\left [ \int \frac {[R_{\alpha \alpha'l_n}(p)+F_{\alpha
\alpha'l_n}(p)]^2} {E_{\alpha}-E_{\alpha '}+p}pdp - \int \frac
{[R_{\beta \alpha'l_n}(p)+F_{\beta \alpha'l_n}(p)]^2}
{E_{\beta}+E_{\alpha '}+p}pdp \right ]
\]
%\begin{eqnarray}
\[
 -\left [{\cal
B}^{jj'm_jm_j'}_{LL'L^*L'^*l_n} -{\cal
D}^{jj'm_jm_j'}_{LL'L^*L'^*l_n} + 2{\cal
E}^{jj'm_jm_j'}_{LL'L^*L'^*l_n} \right ]
\]
\begin{equation}
 \left[ \int\frac{dp \,p}{E_{\alpha}-E_{\alpha '}+p}
{\cal H}_{\alpha \alpha' l_n L L'L^*L'^*}
 - \int\frac{dp\,p}{E_{\beta}+E_{\alpha '}+p}
{\cal H}_{\beta \alpha'l_nLL'L^*L'^*}
 \right ] \biggr \} ,
\end{equation}
%\]
where we define function
\[
{\cal H}_{\alpha\alpha'l_nLL'L^*L'^*}= {\cal
H}_{\alpha\alpha'l_nLL'L^*L'^*}(p)
\]
\begin{equation}
= H^2_{\alpha \alpha 'l_n}\delta_{lLL^*}\delta_{l'^{\pm}L'L'^*} +
H^2_{\alpha ' \alpha l_n}\delta_{l^{\pm}LL^*}\delta_{l'L'L'^*} -
H_{\alpha \alpha 'l_n} H_{\alpha ' \alpha l_n} (\delta_{lL}
\delta_{l^{\pm}L^*} \delta_{l'^{\pm}L'} \delta_{l'L'^*}
+\delta_{lL^*} \delta_{l^{\pm}L} \delta_{l'L'} \delta_{l'^{\pm}
L'^*} )
\end{equation}
and radial integrals
\[
 H_{\alpha \alpha 'l_n}= H_{\alpha \alpha 'l_n}(p)=
 \int dr\left[r^2 f_{\alpha'}(r)g_{\alpha}(r)j_{l_n}(pr)\right],
\]
\[
R_{\alpha \alpha 'l_n}= R_{\alpha \alpha 'l_n}(p)=
 \int dr\left[r^2 g_{\alpha'}(r)g_{\alpha}(r)j_{l_n}(pr)\right],
\]
\begin{equation}
F_{\alpha \alpha 'l_n}= F_{\alpha \alpha 'l_n}(p)=
 \int dr\left[r^2 f_{\alpha'}(r)f_{\alpha}(r)j_{l_n}(pr)\right].
\end{equation}

The angular momentum coefficients ${\cal A}, \, {\cal B}, \, {\cal
D}$ and ${\cal E}$ are given in the Appendix C. We note also that
the sum in Eq.(26) does not depend on the orientation of the full
momentum of the valence quark $m_j'=-j',-j'+1,...,j'$.

\subsection{ Exchange diagrams contributions}
 \par
\begin{figure}[tbh]
\begin{center}
\includegraphics[width=15cm]{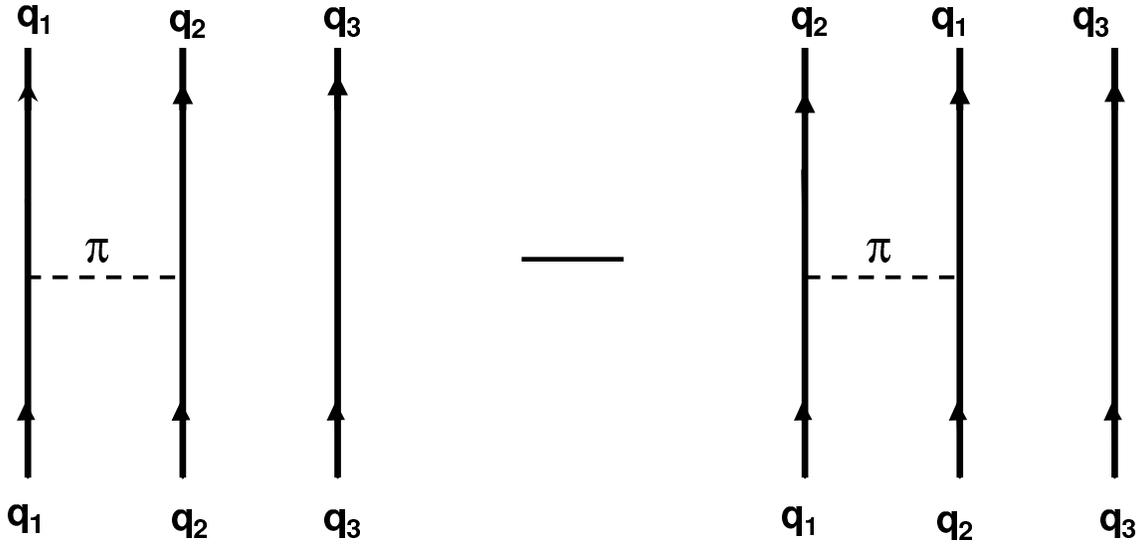}
\end{center}
\caption{Second order $\pi-$meson exchange diagrams
 \label{Fig2a}}
\end{figure}

 The pion exchange contribution to the baryon energy-shift (see Fig.\ref{Fig2a} ) is evaluated as:
 \begin{eqnarray}
\Delta
E_{ex.}^{(\pi)}=-\frac{1}{2f_{\pi}^2}\sum\limits_{a=1}^{3}\sum\limits_{\alpha
 \leq  \alpha_F}\sum\limits_{\alpha ' \leq \alpha_F} \int\frac{d^3\vec p}
{(2\pi)^3p_0^2} \biggl\{ V_{\alpha \alpha  }^{a+}(\vec p)
V_{\alpha '\alpha ' }^{a}(\vec p)- V_{\alpha \alpha '}^{a+}(\vec p)
V_{\alpha  \alpha '}^{a}(\vec p) \biggr\}.
\end{eqnarray}
 By using the Wick's theorem we can write a more convenient  expression for the
 energy shift of the SU(2) baryons from the  second order pion exchange
 diagrams:
  \begin{eqnarray}
\Delta E_{ex.}^{(\pi)}=-\frac{1}{16\pi^3f_{\pi}^2}\int
\frac{d p \, p^2}{p_0^2} \sum\limits_{l_n}\Pi_{l_n}(p)
\end{eqnarray}
where
\begin{eqnarray}
\label{pionex}
 \Pi_{l_n}(p)=<\Phi_B|\sum\limits_{i\neq j}\vec
\tau(i)\vec\tau(j)T_{ln}(i) T_{l_n}(j)K_{l_n}(i)K_{l_n}^+(j)|\Phi_B>
\,
\end{eqnarray}
and the operators $\vec \tau, T_{l_n}$ and $ K_{l_n}$ are summed
over single quark levels $i\neq j$ of the SU(2) baryon. In the quark
model, the baryon wave function $ |\Phi_B> $ is presented as a bound
state of three valence quarks, and it can be written down commonly
as
\begin{eqnarray}
\nonumber |\Phi_B>=|\alpha\beta\gamma>=\sum
\limits_{J_0T_0}|\alpha\beta;\gamma>_{JM(J_0)}^{TM_T(T_0)} \\
\nonumber =\sum\limits_{J_0T_0}\hat {S}\biggl [
|\psi_{\alpha}(r_1)\psi_{\beta}(r_2) \psi_{\gamma}(r_3){\cal
Y}_{J_0}^{JM}(\hat{x_1}\hat{x_2};\hat{x_3})>
|\chi_{T_0}^{TM_T}(12;3)>\biggr ]|\chi_c^(123)>,
\end{eqnarray}
where $J_0$ and $T_0$ are intermediate spin and isospin couplings,
respectively. The states $\psi$ are the single particle states,
labeled by a set of quantum numbers $\alpha$, $\beta$ and $\gamma$,
excluding the color  degree of freedom.

\par  The operator $T_{l_n}$ in equation (\ref{pionex}) is the
radial integration operator:
\begin{eqnarray}
<\alpha|T_{l_n}|\beta>=\int dr\biggl [r^2
S(r)j_{l_n}(pr)G_{\alpha\beta}(r)\biggr ].
\end{eqnarray}
with
%$\alpha=(N,l,j,m_j,m_t,m_c)$ and $\alpha^{\prime}$ is defined as
\begin{eqnarray}
  G_{\alpha \alpha '}(r)=f_{\alpha}(r)g_{\alpha '}(r) + f_{\alpha '}(r)
g_{\alpha }(r)  .
\end{eqnarray}
 where $\alpha=(N,l,j,m_j,m_t,m_c)$ and $\alpha^{\prime}$ are
 two sets of the single quark quantum numbers.
 The matrix elements of the operator $K_{l_n}$ are given by
\begin{eqnarray}
\nonumber
<\alpha|K_{l_n}|\beta>=-\biggl( 4\pi
[l^{\pm}(\alpha)][l_n][j(\alpha)]\biggr)^{1/2}C^{l(\beta)0}_{l^{\pm}
(\alpha)0l_n 0} \\
\nonumber
W (j(\alpha)\frac{1}{2}l_n,l(\beta); l^{\pm(\alpha)},j(\beta))
C^{j(\beta)m(\beta)}_{j(\alpha)m_j(\alpha)l_n(m(\beta)-m(\alpha))},
\nonumber
\end{eqnarray}
and the Hermitian conjunction
\begin{equation}
<\alpha|K_{l_n}^+|\beta>=<\beta|K_{l_n}|\alpha>,
\end{equation}
where $j(\alpha), l(\alpha), l^{\pm}(\alpha), m(\alpha)$ are the quantum
numbers of the single quark state $<\alpha|$.

\par
\begin{figure}[tbh]
\begin{center}
\includegraphics[width=15cm]{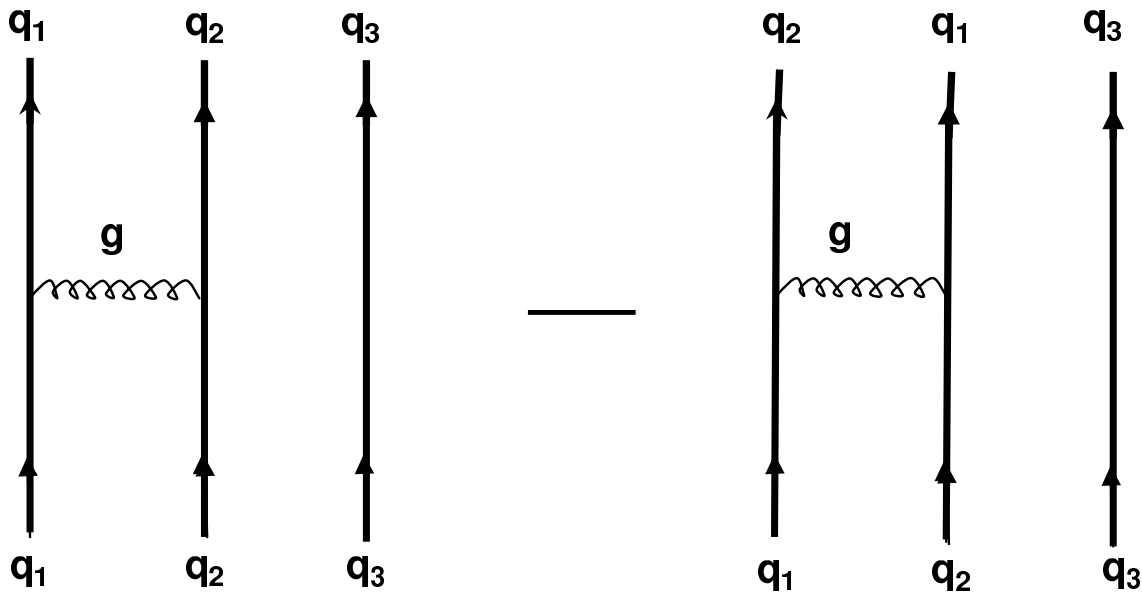}
\end{center}
\caption{Second order gluon-exchange diagrams
 \label{Fig2b}}
\end{figure}

 The contribution of the second-order gluon-exchange terms to the baryon spectrum
 (see Fig.\ref{Fig2b}) is given by
 \begin{eqnarray}
\Delta
E_{ex.}^{(g)}=-\frac{g^2}{2}\sum\limits_{a \mu \nu}\sum\limits_{\alpha
  \leq \alpha_F}\sum\limits_{\alpha ' \leq \alpha_F} \int\frac{d^3\vec p}
{(2\pi)^3p^2} \biggl \{ V_{\alpha \alpha  }^{a\mu+}(\vec p)
V_{\alpha '\alpha ' }^{a\nu}(\vec p)- V_{\alpha \alpha '}^{a\mu+}(\vec p)
V_{\alpha  \alpha '}^{a\nu}(\vec p) \biggr \} g^{\mu\nu}.
\end{eqnarray}
By using the Wick's theorem we can write  more convenient expression
for this equation
  \begin{eqnarray}
\Delta E_{ex.}^{(g)}=-\frac{g^2}{\pi}\int \limits_0^{\infty}dp
 \sum\limits_{l_n m_n} {\cal Q}_{l_n m_n}(p)
\end{eqnarray}
with the corresponding color-electric (Coulomb) and color-magnetic
parts:
\begin{eqnarray}
\label{gluon}
 \nonumber {\cal Q}_{l_n m_n}(p)
=<\Phi_B|\sum\limits_{i \neq j}\frac{\vec \lambda(i)}{2} \frac{\vec
\lambda(j)}{2}T^{(g)}_{ln}(i)T^{(g)}_{ln}(j)
\hat F_{l_n m_n}(i) \hat F_{l_n m_n}^+(j) |\Phi_B> \,  \\
%\nonumber
-<\Phi_B|\sum\limits_{i \neq j}\frac{\vec \lambda(i)}{2}
\frac{\vec \lambda(j)}{2}T^{(g)}_{ln}(i)T^{(g)}_{ln}(j)
\hat F_{l_n m_n}(i) \hat F_{l_n m_n}^+(j) \hat {\vec \alpha}(i)
\hat {\vec \alpha}(j)|\Phi_B> .
\end{eqnarray}

 The operator $T^{(g)}_{ln}$ is the radial integration operator with the
 factor $j_{l_n}(pr)$. The operators $\hat F_{l_n m_n}(i)$
and $\hat F^+_{l_n m_n}(j)$ are the angular integration operator
with the factors
 $Y_{l_nm_n}(\hat x_i)$ and $Y^*_{l_nm_n}(\hat x_j)$ respectively. All these
 operators are summed over single quark levels $i \neq j$ of the SU(2) baryon.
\par We note that the relativistic one-pion and one-gluon
(color-magnetic) exchange forces have strongly different structure
from corresponding non-relativistic exchange forces: the exchanged
pion or gluon in the $l_n$-orbit couple the upper (lower) component
of i-valence quark with the lower (upper) component of another
j-valence quark (see above equations) and generate  spin 0, 1, 2,
etc. operators. As we see below, the only contribution to the g.s. N
and $\Delta$ comes from the pion exchange and gluon exchange
(color-magnetic part) via the spin 1 operator (an analog of the
spin-orbit operator), while in the non-relativistic case the
corresponding terms are spin 0 (spin-spin) operators. Another
important point is that the both exchange operators are symmetric
under the permutation of the upper and the lower components of the
interacting valence quarks in i- and j-orbits. This fact indicates
the chiral invariance of the Lagrangian in the matrix elements
level.

\section{Numerical results}
%\subsection{Ground state $N$ and $\Delta$}
\par In order to account for the finite size effect of the pion, we introduce
a one-pion vertex regularization function in the momentum space,
parameterized in the dipole form as
$$ F_{\pi}(p^2)= \frac {\Lambda_{\pi}^2-m_{\pi}^2}{\Lambda_{\pi}^2+p^2}.$$
From the flux-tube study \cite{lus81} we fix $\alpha=0.26 \approx
\pi/12$. The strong coupling constant $g_s^2=4\pi\alpha_s$ with the
value $\alpha_s=0.65$. The parameters of the confining potential
($c$ and  $m$) are chosen to reproduce the correct axial charge
$g_A$ of the proton
 (and the empirical pion-nucleon coupling constant $G^2_{\pi NN}/4\pi=14$ via the
 Goldberger-Treiman relation) and a normal value for the quark core RMS
 radius of the proton (see \cite{gutsche87}).
\par In Table 1 we indicate the model parameters together with the corresponding
single-quark energies for the models $A$ and $B$. By examining the
two model parameters we can check the sensitivity of the baryon
spectrum on the description of the static properties of the proton.
\par Table 2 contains the quark core results \cite{gutsche87} for the
static properties of the proton with the correction on the center of
mass motion (CM). One can note from the table that a larger value of
the strength parameter $c$ (model $B$) of the confining potential
yields a smaller value for the proton RMS radius. Further we will
see that the calculated spectrum is quite sensitive on the choice of
the the strength parameter, hence on the proton RMS  radius.

\par In Table 3 we give the perturbative corrections to the  valence quarks
energy shifts in the $1S, 2S, 1P_{1/2}, 1P_{3/2}$ orbits from the
self-energy diagrams induced by pion and gluon fields. Contributions
from intermediate quark and anti-quark states are included with the
full momentum up to $j=25/2$ which enables  convergent  results
opposite to the bag models (see \cite{gutsche89}). The columns with
$I=0$ corresponds to the special cases where an intermediate quark
is in the same orbit as the initial and final quark. These cases are
useful to check the limitation of the self energy in the literature
with the intermediate ground state of the baryon. As one can note
from the table, the self-energy results with the restriction $I=0$
are negative for the both pion and gluon (color-magnetic part)
fields, while are positive without the restriction (due to the
intermediate anti-quark states). Thus the restriction $I=0$
decreases the baryon energy contrary to the complete sum of
intermediate quark and anti-quark states.
\par The contributions of the exchange diagrams to the energy-shift
of the lower SU(2) baryon states are given in Tables 4 and 5. As was
pointed out above, the relativistic structures of the pion and gluon
exchange mechanisms are different from the corresponding
non-relativistic operators. In first case the both pion and gluon
exchange operators couple the upper (lower) and lower (upper) Dirac
components of the two interacting valence quarks. One can see from
Table 4 that in the ground and radially excited states of the $N$
and $\Delta$ where all three valence quarks are in the $S$-orbits,
the exchanged pion or gluon can be only in the $l_n=1$ orbit, i.e.
generate spin 1 operators (analogy of the spin-orbit operator).
These operators, naturally, can contribute to the proton spin like
the lower component (p-wave) of the Dirac wave function. The
non-relativistic pion- and gluon-exchange operators between valence
quarks in the ground state have a spin-spin structure, i.e. are spin
0 operators and as a result do not contribute to the proton spin.
 For the baryon states
with the last excited quark in the $1P_{1/2}$-orbit, the
contribution comes from the both $l_n=0$ (spin-spin) and $l_n=1$
(spin-orbit) terms. For baryons with a structure $(1S)^2 1P_{3/2}$
the corresponding exchange forces generate the $l_n=1$ (spin-orbit)
and $l_n=2$ (tensor) operators. From the Table 5 one can note that
the color-electric forces yield large spin-independent and small
spin-dependent matrix elements. In the case when the intermediate
quark states are restricted to the ground state in the $N(939)$ and
$\Delta(1232)$, the color-electric component of the gluon-cloud
contribution is exactly canceled by the corresponding exchange term
in consistence with the MIT bag-model result.
\par The analysis of the results presented for the excited baryon states
with the structure $(1S)^21P_{1/2}$ and $(1S)^21P_{3/2}$ indicates
that the spin-orbit matrix elements generated by the both pion- and
gluon-exchange (color-magnetic) forces are of acceptable order
contrary to the non-relativistic one-gluon exchange spin-orbit
matrix elements \cite{isg77,isg78}.

\par In Table 6 we give the mass values for the g.s. N(939) with and
without CM correction for the both Model A and B. The restriction to
the intermediate ground state when estimating the self-energy (I=0)
yields too small values for the nucleon energy (and also for other
baryons). Contrary, the inclusion of all excited intermediate states
results large mass values.  Model B yields even larger value for the
mass of the nucleon. This means that larger values of the strength
parameter $c$ of the confining potential is not likely. Another
important point is that the inclusion of the gluon-exchange and
self-energy diagrams do not improve the situation. For the model A,
the contribution of the pion fields to the $N(939)$ is about 200
MeV, while the gluon fields yield more than 500 MeV shift with the
strong coupling constant $\alpha_s=0.65$.
\par One can note that the three methods for the
correction of center of mass motion agree within 50 MeV which seems
too large. However, these three methods always give corrections with
systematic differences. For example, the LHO method yields
correction larger than the $P=0$ method, but smaller than the $R=0$
method. Thus, we can fix one of these methods and go to the excited
sector.
\par And finally in last Table 7 we compare
the theoretical energy values including the CM correction in the LHO
method  with experimental data from PDG \cite{PDG06}. We assume that
the lower negative parity excited N and $\Delta$ states correspond
to the excited last valence quark in the $1P_{3/2}$-orbit, while the
radially excited Nucleon $N^*(1440)$
 (Roper ) and Delta $\Delta^*(1600)$ resonances are assigned with the
 radially (in 2S-state) excited valence quark.
\par We note, that one can recalculate the whole gluon-loop corrections to  the
SU(2) baryon spectrum by changing the value of the strong coupling
constant (we use $\alpha_s=0.65$) in order to have smaller
contribution to the mass-shift values. However, this way yields
smaller values for the splittings  between SU(2) baryon states. The
situation is not helpful for the $N-\Delta$ splitting, since we have
only 63 MeV with the above value of the strong coupling constant.
Thus, we come to the conclusion that the gluon field corrections to
the mass-shift values of the SU(2) baryon states must be small and
confirm the results of the non-relativistic CQM results
\cite{glo97}.
 \par When ignoring the gluon loop corrections, we have an overall
 good description of the SU(2) baryon spectrum. The Nucleon g.s.
 and the Roper resonance $N^*(1440)(1/2^+)$ are overestimated by more than 200 MeV,
 while the orbital excitations of the Nucleon, $N^*(1520)(3/2^-)$ and
 $N^*(1535)(1/2^-)$  are slightly underestimated by about 40-60 MeV.
  \par The situation in the Delta sector is quite different.
  Unlike the Nucleon sector, the $\Delta(1232)(3/2+)$ resonance and
  it's first radial excitation $\Delta(1600)(3/2+)$ are overestimated slightly
  by 50-90 MeV in the Model A. But, the orbitally excited  Delta
resonances $\Delta(1620)(1/2-)$ and $\Delta(1700)(3/2-)$ are also
underestimated by small values (like the Nucleon sector).
 \par   In order to check whether the gluon field contributions can be replaced by the
 multiple pion exchange mechanism, we repeat the calculations of the ground state Nucleon
 and $\Delta(1232)(3/2+)$ with smaller and larger values of the pion
 decay constant. For the $f_{\pi}^{\prime}=$ 1.5 $f_{\pi}$ we obtained
 $m(N)=1055$ MeV and $m(\Delta)=$1118 MeV which have to be compared
 with the numbers 1166 MeV and 1310 MeV from the Table 7, respectively. For the
 value $f_{\pi}^{\prime}=$ 0.5 $f_{\pi}$ we have $m(N)=1766$ MeV and
 $m(\Delta)=$2342 MeV, which are very large comparing the corresponding
 estimations calculated with the experimental pion decay constant $f_{\pi}$.

 \par The different level of the estimations for the Nucleon and Delta
 sectors indicates on the necessity of the strong Instanton induced exchange
 mechanism, which does not change the Delta spectrum, while
 effecting the Nucleons essentially. Additionally, the two-pion loop
 corrections are needed for the both sector, which would yield
 small corrections.

\section{Summary and conclusions}
\par We have developed a relativistic chiral quark model with the
inclusion of the gluon and pion fields contribution to the mass
spectrum of the SU(2) baryons.  These fields were shown to generate
the second-order self-energy and exchange corrections. The overall
good description of the lower excitation spectrum of the $N$ and
$\Delta$ states are obtained when restricting to the pion fields
corrections. The color-electric components of the gluon fields give
enormously large contributions to the baryon spectrum (about 500 MeV
with $\alpha_s=0.65$ for the nucleon g.s.).
 The exchanged pion or gluon (color-magnetic component) generates the
spin 0 (spin-spin), spin 1 (spin-orbit) and spin 2 (tensor)
operators with reasonable values of the matrix elements for the
SU(2) baryons energy shifts, thus indicating no any "spin-orbit
problem" in the excited Nucleon and Delta sectors. The only
contribution to the g.s. Nucleon and Delta comes from the spin 1
(spin-orbit) operator which can contribute also to the baryon spin.

\par When using a small value of the strong coupling constant or
completely ignoring the gluon loop corrections,  the model yields a
quite good description of the Delta resonances mass-spectrum
 within 50-60 MeV. However, the calculated Nucleon sector differs from the experimental
 values essentially.

\par In conclusion, the developed model still needs an additional mechanism
for the description of the excited Nucleon and $\Delta$ spectrum.
Possible candidates are the two-pion exchange and the Instanton
induced exchange (for the nucleon sector) forces.

{\it Acknowledgements}. The work is supported in part by the DAAD
Research fellowship (Germany) and by the Scientific Fund for the
Support of Fundamental Research, Uzbekistan Academy of Sciences. The
author acknowledges the Institut f\"ur Theoretische Physik,
Universit\"at T\"ubingen for kind hospitality and thanks D.Baye,
A.Faessler, Th. Gutsche, A. Rakhimov, F. Stancu, C. Semay for useful
discussions.

\clearpage
\appendix
\begin{center}
\large\bf {Appendix A: $\pi$-q-q transition form factors}
\end{center}
\vskip 0.5cm
\par Putting explicit expression of the vertex matrix
$\Gamma^a(\vec x)$ from Eq.(19) into Eq.(17) we receive next
equation:
\begin{eqnarray}
\nonumber V_{\alpha\alpha'}^a(\vec p)&=& -i \int dr
r^2\Big[g_{\alpha}(r)f_{\alpha'}(r)+
g_{\alpha'}(r)f_{\alpha}(r)\Big]S(r) \\
&   &   \int d\hat{r}\Big[ {\cal
Y}_{jl}^{m_j^+}(\hat{r})(\vec{\sigma}\hat{r}) {\cal
Y}_{j'l'}^{m_j'}(\hat{r})e^{-i\vec{p}\vec{r}}\Big]
<m_t|\tau^a|m_t'><m_c|I_c|m_c'>
\end{eqnarray}
Now using
$${\cal Y}_{jl}^{m_j^+}(\hat{r})(\vec{\sigma}\hat{r})=-{\cal Y}_{jl^{\pm}}^{m_j^+}(\hat{r})
$$
which couples the lower orbital momentum to the spin, and expanding
the exponential function over spherical Bessel functions and
integrating over angular part of the variable $\vec r$, we get next
equation for the integral
$$\int d\hat{r}\Big[ {\cal
Y}_{jl}^{m_j^+}(\hat{r})(\vec{\sigma}\hat{r}) {\cal
Y}_{j'l'}^{m_j'}(\hat{r})e^{-i\vec{p}\vec{r}}\Big]=
\sum_{l_n}(-i)^{l_n} j_{l_n}(pr) Y_{l_n}^{m_j'-m_j}(\hat{p})
 {\cal F}(l^{\pm},l',l_n,j,j',m_j,m_j'),
 $$
 where coefficients ${\cal F}$ are defined as
 $$ {\cal F}(l^{\pm},l',l_n,j,j',m_j,m_j')=
-\biggl( 4\pi [l^{\pm}][l_n][j]\biggr)^{1/2}C^{l'0}_{l^{\pm}0l_n0}
  W (j\frac{1}{2}l_n,l'; l^{\pm},j')
C^{j'm'_j}_{jm_jl_n(m'_j-m_j)}.
$$
For the transition form-factor now it is easy to write the next
expression:
\begin{eqnarray}
\nonumber V_{\alpha\alpha'}^a(\vec p) & = & \sum_{l_n}
(-i)^{l_n+1}\int dr r^2\Big[g_{\alpha}(r)f_{\alpha'}(r)+
g_{\alpha'}(r)f_{\alpha}(r)\Big]S(r) j_{l_n}(pr) \\
 &  &  Y_{l_n}^{m_j'-m_j}(\hat{p}) {\cal F}(l^{\pm},l',l_n,j,j',m_j,m_j')
<m_t|\tau^a|m_t'><m_c|I_c|m_c'>.
\end{eqnarray}
The Hermitian conjunction of the transition form factor
\begin{eqnarray}
\nonumber V_{\alpha\alpha'}^{a+}(\vec p)& = & \sum_{l_n}
(i)^{l_n+1}\int dr r^2\Big[g_{\alpha}(r)f_{\alpha'}(r)+
g_{\alpha'}(r)f_{\alpha}(r)\Big]S(r) j_{l_n}(pr) \\
 &  &  Y_{l_n}^{(m_j'-m_j)*}(\hat{p}) {\cal
F}(l^{\pm},l',l_n,j,j',m_j,m_j') <m_t'|\tau^a|m_t><m_c'|I_c|m_c>.
\end{eqnarray}

\appendix
\begin{center}
\large\bf {Appendix B: g-q-q transition form factor}
\end{center}

 The gluon-quark-quark transition form-factor is defined as
\begin{displaymath}
V_{\alpha
\alpha'}^{a\mu}(\vec{p})=\int{d^3x\bar{u}_{\alpha}(\vec{x})\Gamma_{\mu}^{\alpha}
u_{\alpha'}(\vec{x})exp({-i\vec{p}\vec{x}})} ,
\end{displaymath}
where the vertex matrix
\begin{displaymath}
\Gamma_{\mu}^{a}\equiv\gamma^{\mu}\frac{\lambda_a}{2}I_t .
\end{displaymath}

 Using the properties of the Dirac matrices
 \begin{equation}
 \gamma^0 \gamma^{\mu} = \delta_{\mu 0}I + \hat{\alpha}_k \delta_{\mu k}
 \end{equation}
one can write:
\begin{equation}
V_{\alpha \alpha'}^{a\mu}(\vec{p})=\delta_{\mu 0}
\int{d^3x\bar{u}_{\alpha}(\vec{x})\frac{\lambda_a}{2}I_t
u_{\alpha'}(\vec{x})exp({-i\vec{p}\vec{x}})} + \delta_{\mu k}
\int{d^3x\bar{u}_{\alpha}(\vec{x})\frac{\lambda_a}{2}I_t
\hat{\alpha}_k u_{\alpha'}(\vec{x})exp({-i\vec{p}\vec{x}})}
\end{equation}
The last expression is convenient for the estimation of the exchange
diagrams.
\par For the self-energy diagrams we use an alternative
expression of the transition form-factors. Putting the quark wave
functions with further integration over the radial part of the
 spatial coordinate one can write for the transition form-factor
 next equation:

\begin{eqnarray}
\nonumber\
V_{\alpha\alpha'}^{a\mu}(\vec{p})=\sum_{l_{n}m_{n}}\sum_{LL'}
\sum_{m_{L}m_{L}'m_{s}m_{s}'}\Big(\frac{[L][l_n](4\pi)}
{[L']}\Big)^{\frac{1}{2}}(-i)^{l_{n}}
Y_{l_{n}m_{n}}(\hat{p})M_{m_{s}m_{s}'}^{\mu}C_{\mathrm{L0l_{n}0}}^{\mathrm{L'0}}
C_{\mathrm{Lm_{L}\frac{1}{2}m_{s}}}^{\mathrm{jm_{j}}}
C_{\mathrm{L'm_{L}'\frac{1}{2}m_{s}'}}^{\mathrm{j'm_{j}'}}
C_{\mathrm{Lm_{L}l_{n}m_{n}}}^{\mathrm{L'm_{L}'}}\cdot \\
%\nonumber\
\cdot\int{r^{2}R_{\mathrm{\mu_{LL'}}}^{\mathrm{\alpha\alpha'}}(r)j_{\mathrm{l_{n}}}(pr)dr}
<m_{t}|I_{t}|m_{t}'><m_{c}|\frac{\lambda_{a}}{2}|m_c'> ,
\end{eqnarray}
 where the spin transition matrices
\begin{displaymath}
 M_{m_s m_s'}^0 = \delta _{m_s m_s'},
\end{displaymath}
and
\begin{displaymath}
 M_{m_s m_s'}^k = \sum _{k' =\pm 1,0} h_{kk'} \Big[
 \delta_{k'1}\delta_{m_s 1/2}\delta_{m_s'(-1/2)} +
\delta_{k'(-1)}\delta_{m_s (-1/2)}\delta_{m_s'1/2} +
 2 m_s \delta_{k'0}\delta_{m_s m_s'} \Big]
\end{displaymath}

with the only nonzero expansion coefficients $
h_{1,+1}=h_{1,-1}=h_{3,0}=1$, and  $ h_{2,+1}= -h_{2,-1}= -i $ .

\par The radial functions are defined as

\begin{displaymath}
 R_{\mu_{LL'}}^{\alpha\alpha'}(r)=
\delta_{\mu,0}\delta_{Ll}\delta_{L'l'}
(g_{\alpha}g_{\alpha'}+f_{\alpha}f_{\alpha'} ) +i \delta_{\mu,k}
(\delta_{Ll}\delta_{L'l'^{\pm}} g_{\alpha} f_{\alpha'} -
\delta_{L'l'}\delta_{Ll^{\pm}} g_{\alpha'} f_{\alpha})
\end{displaymath}

\begin{center}
\large\bf{ Appendix C: Recoupling of the Clebsch-Gordan
coefficients}
\end{center}
\begin{eqnarray}
\nonumber\ {\cal
A}_{\mathrm{LL'L^{*}L'^{*}l_{n}}}^{\mathrm{jj'm_jm'_j}} \equiv
\sum_{m_L,\cdots,  m_{n}}
C_{Lm_L\frac{1}{2}m_s}^{jm_j}C_{L'm'_L\frac{1}{2}m'_s}^{j'm'_j}
C_{LM_Ll_nm_n}^{L'M'_L} C_{L^{*}m_{L^*}\frac{1}{2}m_{s^*}}^{jm_j}
C_{L'^{*}m'_{L^{*}}\frac{1}{2}m'_{s^{*}}}^{j'm'_j}
C_{L^*m_{L^*}l_nm_n}^{L'^{*}m'_{L^{*}}} \cdot  \\
\nonumber
\delta_{m_sm'_s}\delta_{m_{s^*}m'_{s^{*}}}\delta_{lLL^*}\delta_{l'L'L'^{*}}
\end{eqnarray}

The coefficients ${\cal B}, {\cal D}, {\cal E}$ are sums of the
Clebsch-Gordan coefficients with the corresponding delta factors
(instead of the delta factors of the coefficient ${\cal A}$) :
$\delta_{m_s,m'_s}\delta_{m_{s^*},m'_{s^{*}}}\delta_{m_s,m_{s^*}}$ ,
$\delta_{m_s,m'_s}\delta_{m_{s^*},m'_{s^{*}}}\delta_{m_s,-m_s^*}$,
 and $\delta_{m_s',m'_{s^*}}\delta_{m_s,m_{s^*}}\delta_{m_s,-m'_s}$, respectively.

\newpage
\begin{table}
\caption { Parameter sets for the models A and B and corresponding
single quark energies in MeV}

\begin{tabular}{|c|c|c|c|c|c|c|c|c|c|}     \hline
 Model          & c, Gev$^2$ &  m, Gev & $\Lambda_{\pi}$, Gev &
 $\alpha$ &$\alpha_s$
                & $E(1S)$ & $E(2S)$ &  $E(1P_{3/2})$ & $E(1P_{1/2})$       \\    \hline
   A        & 0.16  & 0.06 &1.0 &0.26 & 0.65& 571.7&986.7& 822.8 & 860.7          \\  \hline
   B        & 0.20  & 0.07 & 1.2 & 0.26&0.65 & 641.4 & 1105.71 &922.0 & 964.4     \\  \hline

\end{tabular}
\end{table}

\begin{table}
\caption { Quark core contributions (with CM-correction) to the
  static properties of the proton \cite{gutsche87}}
\begin{tabular}{|c|c|c|c|c|}          \hline
 Model & $g_A$  &$\mu_p, N.M.$  & RMS radius, fm & $G^2_{\pi NN}/(4\pi )$   \\    \hline
 A     & 1.26   & 1.58      &  0.52   & 13.919                    \\      \hline
 B     & 1.26   & 1.41      &  0.47   & 13.984                    \\    \hline
 % exp.  & 1.26   & 2.793     &  0.81   & 14.0                    \\     \hline
 \end{tabular}
\end{table}
\begin{table}
\caption { Second order perturbative corrections from self energy
diagrams induced by pion and gluon fields for the energy shift of
the valence quarks in MeV }
 \begin{tabular}{|c|c|c|c|c|c|c|c|c|}   \hline
  Model         & 1S    &1S, I=0  &2S   &2S, I=0  &$1P_{1/2}$ &$1P_{1/2}$, I=0 &$1P_{3/2}$ &$1P_{3/2}$, I=0 \\ \hline
    $\pi$       & 126.5 &-53.86 &350 &-13.7 & 248 &-11.3  &262 &-37.4     \\
 A, $g(cm)$     & 47    &-31.7  &70  &-6.5  &72   &-15.5  &48  &-27.6  \\
    $g(ce)$     & 289   &147.9  &315 &99.1  &308   &119.4 &307 &119.6    \\  \hline

   $\pi$       &193    &-83.5  &540 &-21.1 &380  &-18    &400 &-58.8  \\
 B, $g(cm)$    &54.5   &-35    &80  &-6.9  &80   &-17.3  &50  &-30.8      \\
   $g(ce)$     &306.5  &165.3  &330 &110.2 &320  &133.4  &320 &133.9      \\ \hline
 \end{tabular}
\end{table}

\begin{table}
\caption { Second order perturbative corrections from one-pion and
one-gluon exchange operators with different $l_n$  for the energy
shift of lowest $N$ and $\Delta$ states in MeV for the Model A}
 \begin{tabular}{|c|c|c|c|c|c|c|c|c|c|c|c|c|c|}    \hline
   &(J,T)&$(\frac{1}{2},\frac{1}{2})$&$(\frac{1}{2},\frac{1}{2})$&$(\frac{3}{2},\frac{1}{2})$
 &$(\frac{3}{2},\frac{1}{2})$&$(\frac{5}{2},\frac{1}{2})$&$(\frac{5}{2},\frac{1}{2})$&
 $(\frac{1}{2},\frac{3}{2})$&$(\frac{1}{2},\frac{3}{2})$&$(\frac{3}{2},\frac{3}{2})$&
 $(\frac{3}{2},\frac{3}{2})$&$(\frac{5}{2},\frac{3}{2})$&$(\frac{5}{2},\frac{3}{2}) $    \\
  &  &$\pi$ & g(cm)  &$\pi$ & g(cm) &$ \pi$ & g(cm) &$\pi$ & g(cm) &$ \pi$ & g(cm) &$\pi$ & g(cm)  \\    \hline
 $(1S)^3$   &$l_n=1$ & -179.5 &-31.6 &   &    &  &  & & & -35.9 & 31.6 & &  \\   \hline
 $(1S)^22S$ &$l_n=1$ & -120.35&-14.8 &   &    &  &  & & & -24   & 34.4 & &   \\  \hline
 $(1S)^21P_{1/2}$&$l_n=0$ &15.11&-24.95&-37.78&-12.47 & & &7.56   &62.37 &-15.11 &24.95 & & \\
    &$l_n=1$ &-11.42&-24.62&-31.32& 24.62 & & &-31.32&-17.60 &-2.28 &24.61 & & \\  \hline
 $(1S)^21P_{3/2}$&$l_n=1$ &-70.64&-23.46&-110.93&-17.34 &23.21  &30.94& 17.35 &-23.46 &-0.23
    &-3.06 & -29.54 & 30.94 \\
  &$l_n=2$ &-62.14&-7.77 &-39.41 &-5.45  &-12.43 &-2.12& -24.86&15.54  &-17.40 &11.31 & -4.97
    & 4.24 \\  \hline
\end{tabular}
\end{table}

\begin{table}
\caption { Second order perturbative corrections induced by the
color-electric component of the one-gluon exchange operator for the
energy shift of lowest $N$ and $\Delta$ states in MeV for the Model
A}
 \begin{tabular}{|c|c|c|c|c|c|c|c|}    \hline
   &(J,T)&$(\frac{1}{2},\frac{1}{2})$&$(\frac{3}{2},\frac{1}{2})$
 &$(\frac{5}{2},\frac{1}{2})$&$(\frac{1}{2},\frac{3}{2})$&$(\frac{3}{2},\frac{3}{2})$&
 $(\frac{5}{2},\frac{3}{2})$    \\
  &  & &  &  & & &   \\    \hline
 $(1S)^3$   &$l_n=0$ &-443.7 &   &      &       &-443.7  &          \\  \hline
 $(1S)^22S$ &$l_n=0$ &-411.4 &   &      &       &-411.4  &          \\  \hline
 $(1S)^21P_{1/2}$&$l_n=0$ &-410.6&-410.6&       &-410.6  &-410.6 &       \\
                 &$l_n=1$ &7.5   &3.5   &       &-18.2   &-7.0   &       \\  \hline
 $(1S)^21P_{3/2}$&$l_n=0$ &-389.4&-389.4&-389.4 &-389.4  &-389.4 &-389.4  \\
                 &$l_n=1$ &-7.6 &-29.8 &23.4   &15.3   &-7.6   &-46.8        \\  \hline
\end{tabular}
\end{table}

\begin{table}
\caption { The mass value of the g.s. nucleon in MeV  with and
without center of mass (CM) correction}
 \begin{tabular}{|c|c|c|c|c|c|}     \hline
 Model &              & No CM   & R=0, \cite{lu98}& P=0, \cite{teg82}& LHO, \cite{wil89}   \\    \hline
  & $E_Q$                     & 1715    & 940    &985     & 966      \\
  & $E_Q+\Delta E(\pi)$       & 1915    & 1140   &1185    &1166      \\
& $E_Q+\Delta E(\pi), I=0$    & 1374     & 599   &644      & 625     \\
 A& $E_Q+\Delta E(\pi+g)$     & 2447    & 1672   &1717    & 1698      \\
 & $E_Q+\Delta E(\pi+g)$, I=0 & 1247    & 472    & 517    &498     \\ \hline

  & $E_Q$                    & 1924    & 1057  & 1110    & 1088     \\
    & $E_Q+\Delta E(\pi)$    & 2225    & 1358  & 1411    &1389      \\
  & $E_Q+\Delta E(\pi), I=0$ & 1696    & 829   & 882     &860       \\
 B& $E_Q+\Delta E(\pi+g)$    & 2775    & 1908  & 1961    & 1939     \\
  &$E_Q+\Delta E(\pi+g)$, I=0& 1254    & 387   & 440     & 418      \\  \hline
\end{tabular}
\end{table}

\begin{table}
\caption { Estimations for the energy values of the lowest $N$ and
$\Delta$ states in MeV for the Model A with the CM correction in the
LHO method}
 \begin{tabular}{|c|c|c|c|}     \hline
   & $E_Q+\Delta E(\pi)$&  $E_Q+\Delta E(\pi)+\Delta E(g)$ & exp.\cite{PDG06}  \\  \hline

$N(939)(1/2^+)$   & 1166 & 1698  & 938 $\div$ 939    \\  \hline

$N(1440)(1/2^+)$  & 1684  & 2314  &1430 $\div$ 1470  \\  \hline

 $N(1520)(3/2^-)$ & 1473  &2058   &1515 $\div$ 1530   \\  \hline

 $N(1535)(1/2^-)$ & 1490  &2089   &1520 $\div$ 1555   \\ \hline

$ \Delta(1232)(3/2^+)$ & 1310 & 1905  &1230 $\div$ 1234  \\  \hline

$ \Delta(1600)(3/2^+)$ & 1781 &2461  & 1550 $\div$ 1700  \\  \hline

$\Delta(1620)(1/2^-)$ & 1615  &2259  & 1615 $\div$ 1675  \\  \hline
$\Delta(1700)(3/2^-)$ & 1605  &2242  &1670 $\div$ 1770   \\  \hline

\end{tabular}
\end{table}
\end{document}